# Development of a custom on-line ultrasonic vapour analyzer/flowmeter for the ATLAS inner detector, with application to gaseous tracking and Cherenkov detectors.

**R. Bates**[a], **M. Battistin**[b], **S. Berry**[b], **J. Berthoud**[b], **A. Bitadze**[a], **P. Bonneau**[b], **J. Botelho-Direito**[b], **N. Bousson**[c], **G. Boyd**[d], **G. Bozza**[b], **E. Da Riva**[b], **C. Degeorge**[e], **B. DiGirolamo**[b], **M. Doubek**[f], **J. Godlewski**[b], **G. Hallewell**[c], **S. Katunin**[g,*], **D. Lombard**[b], **M. Mathieu**[c], **S. McMahon**[h], **K. Nagai**[i], **E. Perez-Rodriguez**[b], **C. Rossi**[j], **A. Rozanov**[c], **V. Vacek**[f], **M. Vitek**[f] **and L. Zwalinski**[b].

[a] *SUPA School of Physics and Astronomy, University of Glasgow, Glasgow, G62 7QB, UK*
[b] *CERN, 1211 Geneva 23, Switzerland*
[c] *Centre de Physique des Particules de Marseille, 163 Avenue de Luminy, 13288 Marseille, France*
[d] *The Homer L. Dodge Department of Physics and Astronomy, University of Oklahoma, Norman, OK 73019, USA*
[e] *Indiana University Department of Physics, 727 East 3rd St., Bloomington, IN 47405, USA*
[f] *Czech Technical University, Dept. Applied Physics, Technická 4, 166 07 Prague 6, Czech Republic*
[g] *B.P. Konstantinov Petersburg Nuclear Physics Institute (PNPI), 188300 St. Petersburg, Russia*
[h] *Rutherford Appleton Laboratory, Harwell Science and Innovation Campus, Didcot OX11 OQX, UK*
[i] *Graduate School of Pure and Applied Sciences, University of Tsukuba, Ibaraki 305-8577, Japan*
[j] *Department of Mechanical Engineering - Thermal Energy and Air Conditioning Division, Università degli Studi di Genova Via All'Opera Pia 15a - 16145 Genova, Italy*

 *E-mail*: sergey.katunin@cern.ch

ABSTRACT

Precision sound velocity measurements can simultaneously determine binary gas composition and flow. We have developed an analyzer with custom electronics, currently in use in the ATLAS inner detector, with numerous potential applications. The instrument has demonstrated ~0.3% mixture precision for $C_3F_8/C_2F_6$ mixtures and $< 10^{-4}$ resolution for $N_2/C_3F_8$ mixtures. Moderate and high flow versions of the instrument have demonstrated flow resolutions of ± 2% F.S. for flows up to 250 l.min$^{-1}$, and ± 1.9% F.S. for linear flow velocities up to 15 ms$^{-1}$; the latter flow approaching that expected in the vapour return of the thermosiphon fluorocarbon coolant recirculator being built for the ATLAS silicon tracker.

---

[*] Corresponding author.

**Contents**

## 1. Introduction

We describe a combined ultrasonic gas mixture analyzer/flow meter for continuous real time composition analysis of binary gas mixtures. The instrument exploits the phenomenon whereby the sound velocity in a binary gas mixture at known temperature and pressure is a unique function of the molar concentration of the two components of differing molecular weight. The combined flow measurement and mixture analysis algorithm combines sound transit time measurements in opposed directions with measurements of the pressure and temperature of the mixture, and has many applications where knowledge of binary gas composition is required. The molar concentration of the two component vapours is determined from a comparison of on-line sound velocity measurements with velocity-composition look-up table data gathered from prior measurements in calibration mixtures or from theoretical derivations made with an appropriate equation of state.

Ultrasonic binary gas analysis was first used in particle physics for the analysis of the $N_2/C_5F_{12}$ Cherenkov gas radiator of the SLD Cherenkov Ring Imaging Detector [1], and has subsequently been adopted in all the major ring imaging Cherenkov detectors, including DELPHI, COMPASS and LHCb.

The present development is mainly motivated by a possible future upgrade of the ATLAS silicon tracker evaporative cooling system [2,3], in which the currently-used $C_3F_8$ fluorocarbon evaporative coolant will be blended with a more volatile $C_2F_6$ component [3,4] to allow cooling at lower temperatures. Additionally, the present underground compressor-driven $C_3F_8$ circulation plant will be replaced by a thermosiphon [5]. A combined mixture analyzer/flow meter will be installed in the vapour return to the surface condenser, where a flow of ~ 400 $l.s^{-1}$ is expected.

## 2. Principle of operation of the electronics

The electronics of the present instrument is based on a Microchip® dsPIC 16-bit microcontroller with communication to a SCADA computer running a graphical user interface and analysis package written in PVSS-II®. Figure 1 illustrates the major elements of the electronics, which is presently designed to operate with the SensComp Model 600 instrument grade 50 kHz capacitative foil ultrasonic transducer, originally developed during the 1980s for the Polaroid autofocus camera.

The transducer biasing and amplification chain for a single channel is shown in figure 2. The transducer DC bias voltage of around +300V is provided by a TRACO MHV series DC-DC converter. When transmitting, the transducer is excited by a downgoing (~300 ➔ 0 V) square wave pulse generated in a driver circuit from a TTL pulse from the microcontroller. The transducer is AC-coupled to a signal chain containing differential and programmable gain amplifiers followed by a fast comparator.

A fast (40 MHz) transit time clock, generated in the same microcontroller, is started in synchronism with the leading edge of the transmitted 50 kHz sound pulse. The first received sound pulse crossing the user-definable comparator threshold level stops this clock, as shown in figure 3. The time between the transmitted and first received sound pulses is measured by the microcontroller, which also handles communication via a USB/RS232 interface.

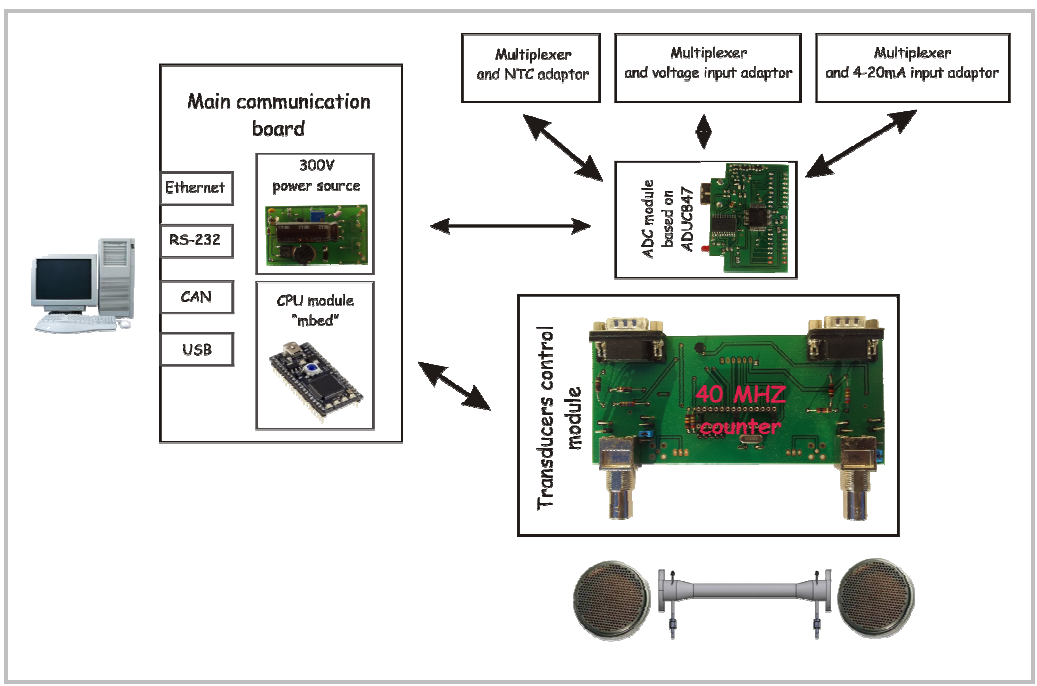

**Figure 1.** Principal elements of the local flowmeter/analyzer electronics.

For flowmetry sound transit times are measured in opposite directions, which may be aligned with the flow of gas or inclined at an angle to it. Rolling average bi-directional transit times, temperature and pressure data continually stream from the FIFO memory to a supervisory computer in which the gas mixture composition and flow rate are continuously calculated using software implemented in PVSS-II version 3.8. When a measuring cycle is requested by the supervisory computer a time-stamped running average from the 300 most recent transit times in each direction in the FIFO memory is output, together with the average temperature and pressure, at a rate of up to 20 averaged samples per second.

In addition to the I/O connectivity for communications, HV for the ultrasonic transducers and analog inputs for pressure and temperature sensors, two (4-20 mA) analog outputs provide feedback for adjustment of the $C_3F_8/C_2F_6$ mixing ratio in an external gas mixture control system.

## 3. Results from different implementations of the instrument

The "pinched axial" implementation of the flowmeter/analyzer - intended for gas flows up to ~ 250 lmin$^{-1}$ - is illustrated in figure 4. Between the transducers, mounted 660 mm apart, vapour flows through a "pinched" tube of inner diameter 44.3 mm, comparable with the transducer diameter of 42.9 mm. Vapour is diverted around the transducers using PEEK® flow-deflecting cones. The temperature and pressure in the tube are monitored to a precision better than ±0.3 °C and ±1 mbar.

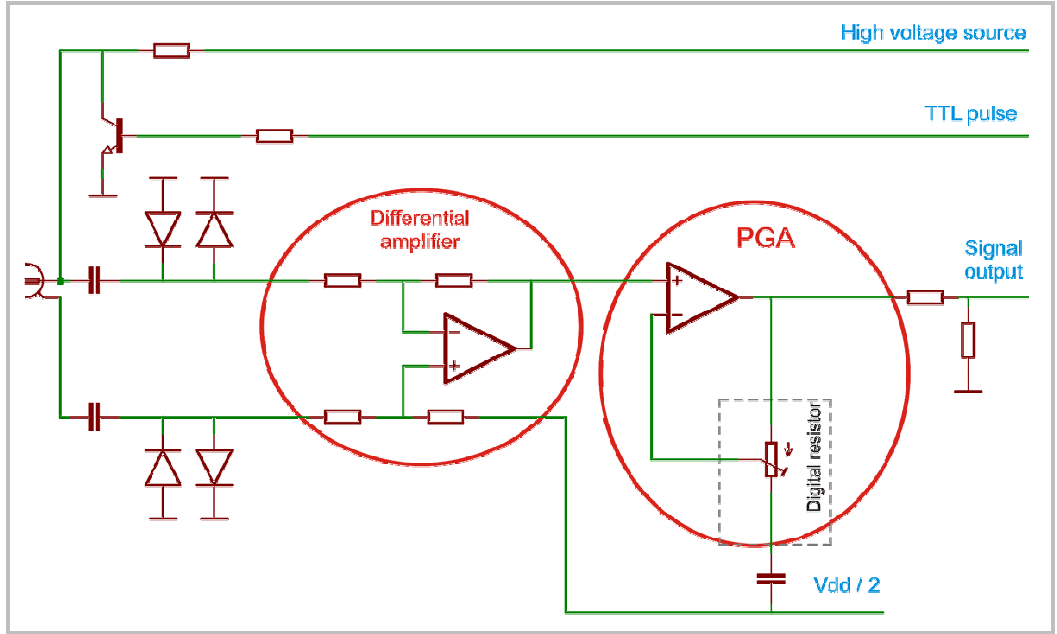

**Figure 2.** Ultrasonic transducer bias and amplification chain (one of two channels).

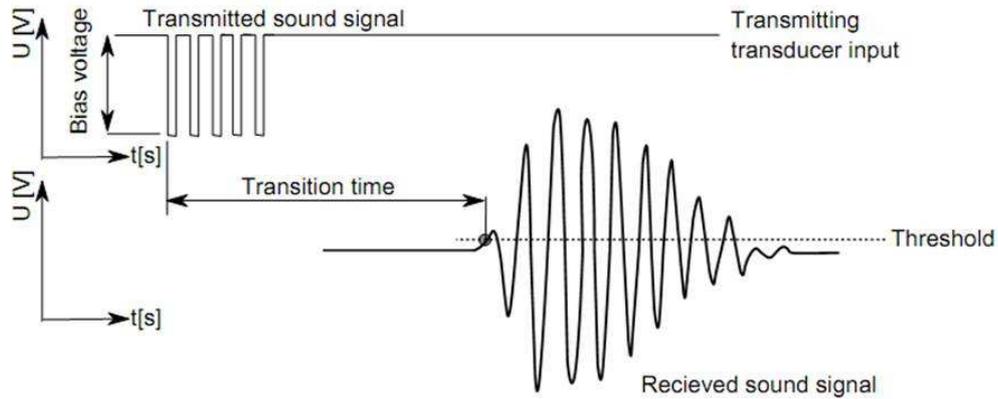

**Figure 3.** Principle of measurement of transit time between the first transmitted sound pulse and the first over-threshold detected pulse.

In this geometry the vapour volume flow rate, $V$ (m$^3$s$^{-1}$), is calculated from the sound transit times measured parallel, $t_{down}$, and anti-parallel, $t_{up}$, to the flow direction, according to:

$$V = \frac{L \cdot A (t_{up} - t_{down})}{2(t_{up} \cdot t_{down})} \qquad (1)$$

where $L$ is the distance (m) between the foils of the transducers and $A$ is the internal cross sectional area (m$^2$) of the tube linking them.

Figure 4 also illustrates the linearity of the ultrasonic flow meter element of the instrument in $C_3F_8$ vapour at 20 °C and 1 bar$_{abs}$ ($C_3F_8$ density ~7.9 kgm$^{-3}$) through comparison with a Schlumberger Delta G16 volumetric gas meter (maximum flow rate 25 m$^3$hr$^{-1}$ [417 lmin$^{-1}$], with precision ± 1 % of full scale – represented as horizontal error bars), at flows up to 230 lmin$^{-1}$; the maximum possible in the presently-available $C_3F_8/C_2F_6$ blend recirculator shown in figure 4.

The vertical error bars reflect the combined uncertainties in the flowmeter tube diameter (± 0.5 mm), transit time measurement precision (± 100ns) and transducer foil spacing (± 0.1 mm after distance calibration in an ideal static gas). The *rms* deviation of the ultrasonic flowmeter relative to the fit (shown as red bands) is ± 4.9 lmin$^{-1}$, around 2 % of the limiting flow of 230 lmin$^{-1}$.

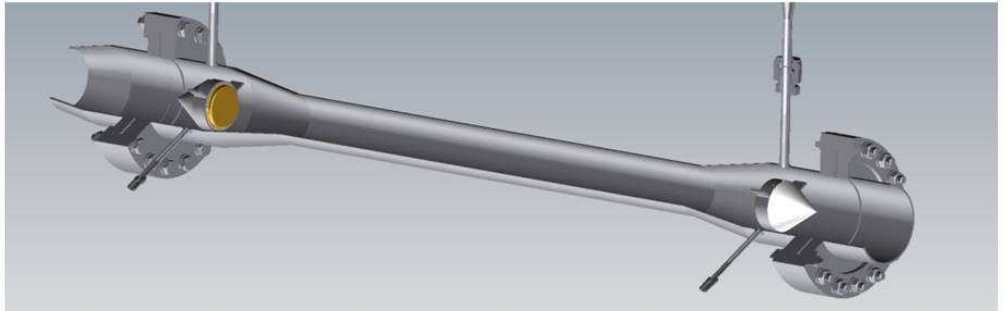

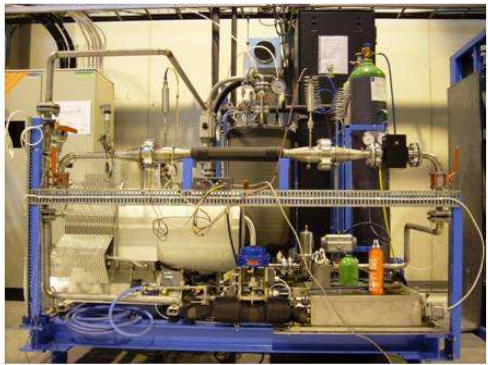 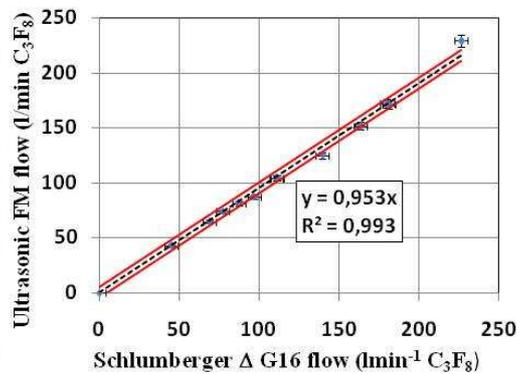

**Figure 4.** "Pinched axial" flowmeter/mixture analyzer: mechanical envelope showing transducer mounting and flow-deflecting cones, together with installation at the $C_2F_6/C_3F_8$ blending and recirculation machine. Linearity comparison is with a Schlumberger Delta G16 gas meter (shown downstream) in $C_3F_8$ at 1 bar$_{abs}$ & 20 °C (density ~7.9 kgm$^{-3}$).

Figure 5 compares the sound velocity measured in the instrument in varying $C_2F_6/C_3F_8$ molar mixing ratios at a temperature of 19.2 °C and a pressure of 1.14 bar$_{abs}$ with predictions from two different theoretical models. The range of mixtures of $C_3F_8$ and $C_2F_6$ spans the region of thermodynamic interest to the ATLAS silicon tracker cooling application.

Contributions to the overall 0.05 ms$^{-1}$ sound velocity measurement error, $\delta c$, were due to:
- ± 0.2 °C temperature stability in the sonar tube (equivalent to ± 0.044 ms$^{-1}$);
- ± 4 mbar pressure stability in the sonar tube (± 0.012 ms$^{-1}$) with the blend circulation machine in operation;
- ± 0.1 mm transducer inter-foil measurement uncertainty (± 0.018 ms$^{-1}$);
- ± 100 ns electronic transit time measurement uncertainty (± 0.002 ms$^{-1}$).

The precision of mixture determination, $\delta(mix)$, at any concentration of the two components is given by;

$$\partial(mix) = \frac{\partial c}{m} \qquad (2)$$

where *m* is the local slope of the sound velocity/ concentration curve (ms$^{-1}$%$^{-1}$). From figure 5 it can be seen that the 0.05 ms$^{-1}$ sound velocity uncertainty yields a concentration uncertainty ~0.3 % at 20 %$C_2F_6$ in $C_3F_8$, where the slope of the velocity/concentration curve is ~ -0.18 ms$^{-1}$%$^{-1}$.

Our sound velocity measurements agree to within ± 0.05ms$^{-1}$ of the predictions of the NIST-REFPROP theoretical package [6], which are shown with ±1% variation in figure 5 for reference.

In a second application related to the ATLAS evaporative cooling system [7], we have been using real-time ultrasonic binary gas analysis for more than a year to detect low levels of $C_3F_8$ vapour leaking into the nitrogen environmental gas surrounding the ATLAS silicon pixel detector.

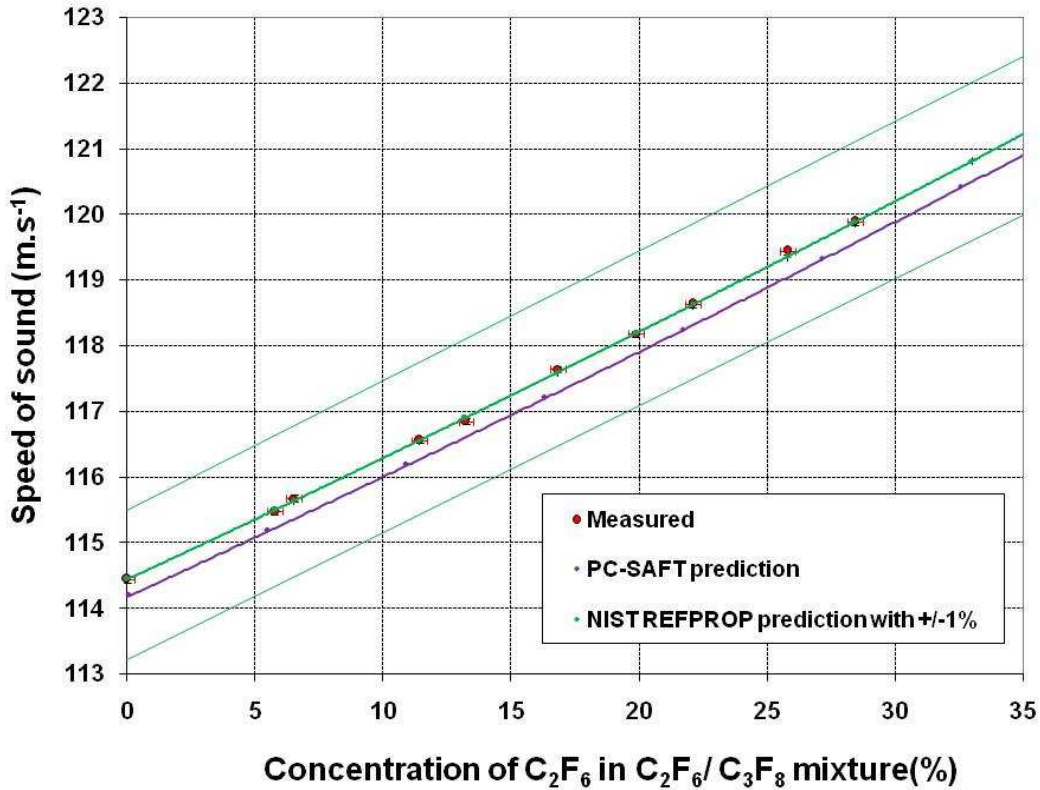

**Figure 5.** Comparison between measured sound velocity data and theoretical predictions in molar $C_3F_8/C_2F_6$ mixtures of thermodynamic interest. NIST-REFPROP predictions shown within a ± 1 % band.

The gas extraction and sampling system is currently being extended (figure 6) to allow measurement of $C_3F_8$ leak rates into the $N_2$ gas envelope of the silicon microstrip (SCT) tracker, and also Xenon leaks from the straw tubes of the TRT (Transition Radiation Tracker) into its external $CO_2$ envelope. For the latter application new electronics is under development for operation in the acoustic frequency range 800-3000Hz: a shift due to the high absorption of sound at ultrasonic frequencies by $CO_2$. This electronics will be reported at a later date.

Under PVSS control gas will be continually aspirated from 7 points (pixel: 1; SCT: 2; TRT: 4) on the ATLAS ID sub-detector environmental gas envelopes and sampled for sequenced analysis via a matrix of normally-open (NO) and normally-closed (NC) pneumatic valves into three simultaneously-operating analysis tubes. Gas exiting the three analysis tubes is vented to an air extraction system for return to the surface. Attachment points for periodic analysis using a portable gas chromatograph (GC) are also available. The sampling rates in the tubes are small enough (<100 $cm^3 min^{-1}$) for the gas to be considered static – flowmetry is of no particular interest in this sampling system.

Figure 7 illustrates a 1-year continuous log of the $C_3F_8$ contamination of the pixel detector environmental $N_2$ envelope. Fluctuations of the measured $C_3F_8$ contamination are correlated with the development of leaks in some of the 88 individual cooling circuits, which have been identified by progressive turn-on or turn-off.

A reduction in sound velocity of ~0.86 $ms^{-1}$ from that of pure nitrogen is typically observed when the full pixel detector cooling system is operating.

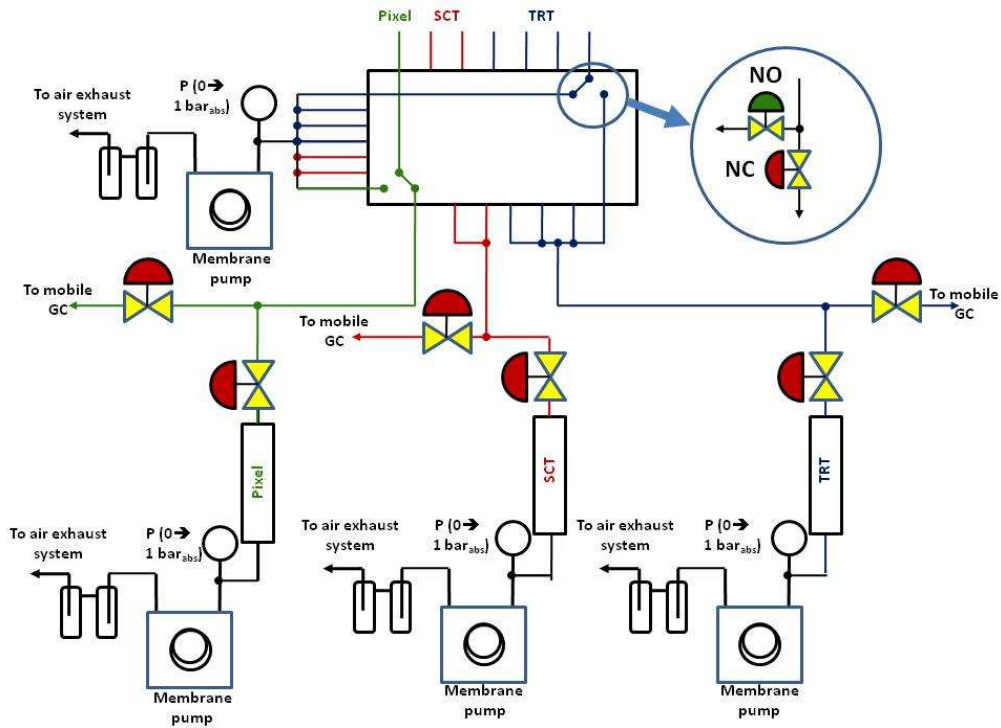

**Figure 6.** Automated sampling system for ultrasonic gas mixture analysis for continuous monitoring of $C_3F_8$ coolant leaks into the $N_2$ volumes surrounding the ATLAS silicon pixel and SCT trackers, and for future acoustic frequency mixture analysis of Xe leaks into the $CO_2$ - containing volume surrounding the ATLAS TRT.

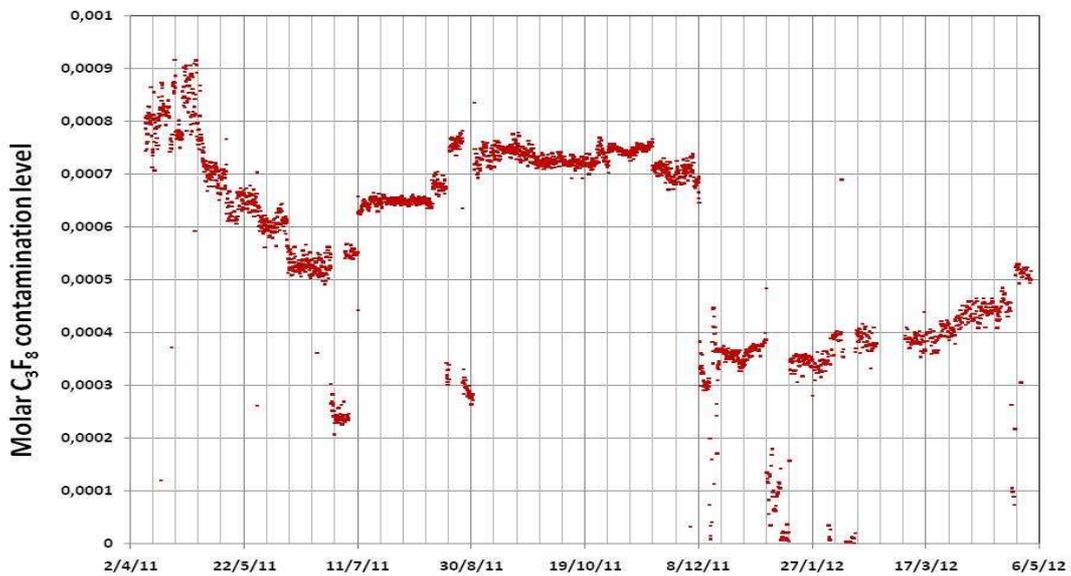

**Figure 7.** Long duration (1 year) log of $C_3F_8$ leak contamination in the $N_2$ environmental gas surrounding the ATLAS pixel detector.

From the ~ -12.27 ms$^{-1}$%$^{-1}$ average gradient of the sound velocity-concentration curve for concentrations in the range 0→0.5 % $C_3F_8$ in $N_2$ this sound velocity difference indicates, via eq. (2), a $C_3F_8$ leak ingress of 0.07 % (figure 7). The intrinsic sound velocity measurement uncertainty of ± 0.05 ms$^{-1}$ correspondingly yields a mixture uncertainty of ± 0.004 % (4.10$^{-5}$).

Following replacement of the present underground compressor-driven $C_3F_8$ circulation plant by a thermosiphon [5], a combined ultrasonic gas mixture analyzer/flow meter will be installed in the single 133 mm diameter vapour return tube to the surface, where fluorocarbon linear flow rates of around 22 ms$^{-1}$ are expected. Computational fluid dynamics (CFD) studies have shown [7] that only an angled crossing geometry with the transducers not impinging on the gas flow is suitable in this application. A prototype angled flowmeter built in PVC tubing with 45º crossing angle was tested - using electronics identical to that used in the pinched axial flowmeter/ analyzer - in comparison with a commercial anemometer, in airflows up to 15 ms$^{-1}$ [7]. The *rms* accuracy of the ultrasonic flowmeter was ± 1.9% of full scale. These positive results have allowed us to construct the final instrument with 45º crossing angle in stainless steel tubing.

## 4. Conclusions and future applications

We have developed a combined real-time flow meter and binary gas analyzer with custom electronics and dedicated SCADA software running under PVSS-II, a CERN standard.

One version of the instrument has demonstrated a resolution of $3.10^{-3}$ for $C_3F_8/C_2F_6$ mixtures with ~20 %$C_2F_6$, and a flow precision of ± 2 % of full scale for fluorocarbon mass flows up to 30 gs$^{-1}$. A sampling instrument [7] has also been in use for more than 1 year to monitor $C_3F_8$ leaks into part of the ATLAS silicon tracker nitrogen envelope. Sensitivity to $C_3F_8$ leak concentrations of $< 10^{-4}$ has been demonstrated in this instrument. The gas sampling system is presently being upgraded to allow two additional analyzers, including an acoustic frequency (1-3 kHz) version for the monitoring of Xenon leaks into the $CO_2$ gas envelope of the ATLAS TRT.

A high flow instrument with a sound path angled at 45° to the gas flow has been developed for the 60 kW thermosiphon recirculator currently under construction for the ATLAS silicon tracker where flow rates of < 400 ls$^{-1}$ (22 ms$^{-1}$) are expected. Tests with a prototype [7] in air at flows up to 15ms$^{-1}$ have demonstrated a flow measurement precision of ± 1.9 % of the full scale. A final instrument is under construction and will be tested in air and with fluorocarbons.

The instruments described in this work have many potential applications where precise and rapid binary gas mixture analysis is required in real-time. Such applications include the analysis and flowmetry of Cherenkov detector radiator gas mixtures, hydrocarbon mixtures, vapour mixtures for semi-conductor manufacture and anaesthetic gas mixtures.